\documentclass[12pt]{iopart}

\usepackage{graphicx} 
\newcommand{\diff}{{\rm d}}
\newcommand{\mathi}{{\rm i}}

\begin{document}
\title[One-dimensional Hubbard model 
in the spin-disordered regime]{Thermodynamics
of the one-dimensional half-filled Hubbard model in the spin-disordered regime}

\author{S Ejima$^1$, F H L Essler$^2$ and F Gebhard$^1$}

\address{$^1$ Fachbereich Physik, Philipps-Universit\"at Marburg,
  Germany\\
$^2$ Rudolf Peierls Centre for Theoretical Physics, Oxford University,
  1 Keble Road, Oxford OX1 3NP, UK}

\ead{florian.gebhard@physik.uni-marburg.de}

\begin{abstract}
We analyze the Thermodynamic Bethe Ansatz equations of the
one-di\-men\-sional half-filled
Hubbard model in the ``spin-disordered regime'', 
which is characterized by
the temperature being much larger than the magnetic energy scale
but small compared to the Mott--Hubbard gap.
In this regime the thermodynamics of the Hubbard model can be thought
of in terms of gapped charged excitations with an effective dispersion
and spin degrees of freedom that only contribute entropically. In
particular, the internal energy and the effective dispersion become
essentially in\-de\-pen\-dent of temperature. An interpretation of this
regime in terms of a putative interacting-electron system at zero
temperature leads to a metal-insulator transition at a finite interaction 
strength above which the gap opens linearly. 
We relate these observations to studies of the Mott--Hubbard transition in the
limit of infinite dimensions.
\end{abstract}

\pacs{71.10.Fd, 71.30.+h}

\vspace{2pc}

\noindent{\it Keywords}: Hubbard model, Thermodynamic Bethe Ansatz,
Mott transition

\vspace{2pc}

%\submitto{\JPA on \today}

\maketitle

\section{Introduction}
\label{sec:intro}

The Mott--Hubbard metal-insulator transition in the one-band Hubbard
model at half band-filling continues to pose an intriguing
problem~\cite{Mott,Gebhardbuch}, even in the limit of infinite
dimensions~\cite{Metzner,RMP}. Starting from the
limit of large interactions, the Mott--Hubbard gap characterizing the
insulating phase diminishes as $U$ is lowered and eventually 
closes at a finite $U_{\rm c}$. 
The determination of the precise value of the latter
is an unsolved problem in infinite dimensions: 
The equations from Dynamical Mean-Field Theory
need to be solved numerically and the best available treatments
using the Dynamical Density-Matrix Renormalization Group method 
lead to conflicting results~\cite{Nishimoto,Uhrig}. 

Other attempts to determine $U_{\rm c}$ involve extrapolations of
finite-order $1/U$-ex\-pan\-sions 
of the Mott--Hubbard gap~\cite{GebhardKalinowski} 
and the ground-state energy~\cite{Bluemer}, respectively.
These expansions can be carried out because the spin
background is completely disordered and one is left with finding the
description of the effective motion of the charge degrees of freedom.
However, the extrapolated values for~$U_{\rm c}$ obtained by such
methods vary significantly, depending on the details of the procedure.

In light of these issues it is desirable to have an example
of a Hubbard-type model with a disordered spin background
which can be solved exactly. 
One such example is the Falicov--Kimball model at half
band-filling in the disordered phase~\cite{FKreview}. However, 
its ground-state energy turns out to have a trivial $1/U$-expansion
and hence is of limited utility.
Curiously, the exactly solvable one-dimensional
Hubbard model~\cite{Zebook} at half band-filling 
features a spin-disordered regime
at a small but finite temperature in the limit of strong Coulomb
repulsion, see Refs.~\cite{Marburger,ChZ,Balents}.
It is then an interesting question to analyze the physics
of this regime in view of the above issues raised by the studies of the
Mott--Hubbard insulator in infinite dimensions.

Our presentation is organized as follows. In section~\ref{sec:TBA}
we review the thermodynamics of the one-dimensional Hubbard
model~\cite{Takahashi7274,MTakahashi,ZHa}. 
In section~\ref{sec:int} we analyze the
Ther\-mo\-dynamic Bethe Ansatz (TBA) equations in the spin-disordered
regime and calculate the internal energy and the effective dispersion
of the charge degrees of freedom. In section~\ref{sec:interpretation}
we interpret these results in terms of a putative 
interacting-electron system at zero temperature. We conclude in 
section~\ref{sec:conclusions}.

\section{Thermodynamics of the one-dimensional Hubbard model}
\label{sec:TBA}

\subsection{Hubbard model and TBA equations}

The Hubbard model on~$L$ lattice sites is described by the Hamiltonian
\begin{equation}
\hat{H}=-t\sum_{j=1}^{L}\sum_{\sigma=\uparrow,\downarrow}
          (\hat{c}_{j,\sigma}^{+} \hat{c}_{j+1,\sigma}^{\phantom{+}}
           +\hat{c}_{j+1,\sigma}^{+} \hat{c}_{j,\sigma}^{\phantom{+}})
+U\sum_{j=1}^{L}(\hat{n}_{j,\uparrow}-1/2)(\hat{n}_{j,\downarrow}-1/2)\; .
\label{eqn:Hamiltonian}
\end{equation}
In the following we define $u=U/(4t)$ and set $t=1$ as our energy unit.
Periodic boundary conditions apply, and the kinetic energy is diagonal
in momentum space with the bare dispersion $\epsilon(k)=-2\cos k$.

The `Thermodynamic Bethe Ansatz (TBA) equations' describe the ratios
of the distributions of hole and particle excitations in thermal
equilibrium at temperature~$T$, chemical potential~$\mu$ and magnetic
field~$B$ (\cite{Takahashi7274}, eq.~(5.60)-(5.65) of~\cite{Zebook})
\begin{eqnarray}
\ln\zeta(k)&=&-\frac{2\cos k}{T}
-\frac{4}{T} \int_{-\infty}^{\infty}\diff\Lambda\ s(\sin k-\Lambda)
{\rm Re}\left[\sqrt{1-(\Lambda-\mathi u)^2}\right] \nonumber \\
&&   + \int_{-\infty}^{\infty}\diff\Lambda\ s(\sin k-\Lambda)
\ln\left(\frac{1+\eta^{\prime}_1(\Lambda)}{1+\eta_1(\Lambda)}\right)\; ,
\label{eqn:TBA2satoshi} \\
\ln \eta_1(\Lambda) &=& 
\left. s*\ln(1+\eta_{2})\right|_{\Lambda} 
-\int_{-\pi}^{\pi}\diff k \cos(k) s(\Lambda-\sin k)\ln[1+1/\zeta(k)]\; ,
\nonumber\\[3pt]
\ln \eta_n(\Lambda) &=& 
\left. s*\ln(1+\eta_{n+1})(1+\eta_{n-1})\right|_{\Lambda} 
\qquad \hbox{for $n\geq 2$}\; ,
\label{simplifiedIE}
\end{eqnarray}
and likewise for $\eta_n^{\prime}(\Lambda)$ with $1/\zeta(k)$ replaced by
$\zeta(k)$. The convolution operation is defined by
\begin{eqnarray}
\left. s*f\right|_x&=&
\int_{-\infty}^{\infty} \diff y\ s(x-y) f(y) \; ,\nonumber\\
s(x) &=& \frac{1}{4u\cosh(\pi x/(2u))} = \int_{-\infty}^{\infty} 
\frac{\diff\omega}{2\pi} 
\frac{\exp\left(-\mathi \omega x\right)}{2\cosh(\omega u)}\; .
\end{eqnarray}
The ``boundary conditions'' are
\begin{equation}
\lim_{n\to\infty}\frac{\eta_n(\Lambda)}{n}=\frac{2B}{T} \quad , \quad
\lim_{n\to\infty}\frac{\eta_n^{\prime}(\Lambda)}{n}=-\frac{2\mu}{T} \; .
\label{boundarycond}
\end{equation}

\subsection{Free-energy density and internal energy}

In terms of the distribution functions $\zeta(k)$ and $\eta_1(\Lambda)$
the free-energy density can be cast into the form (see eq.~(5.69) 
of~\cite{Zebook})
\begin{equation}
  f(T)= e_0-\mu-T\int_{-\pi}^{\pi}\diff k \rho_0(k) \ln(1+\zeta(k))
-T\int_{-\infty}^{\infty}\diff\Lambda\sigma_0(\Lambda)\ln(1+\eta_1(\Lambda))
\, ,
\label{eqn:gibbs2}
\end{equation}
where $e_0$
denotes the ground-state energy of the one-dimensional Hubbard model
at half band-filling, see eq.~(\ref{eqn:ezerozero}).
Here, the root densities 
$\sigma_0(\Lambda)$ and $\rho_0(k)$ are given by 
\begin{eqnarray}
 \sigma_0(\Lambda)&=& \int_{-\infty}^{\infty} \frac{\diff\omega}{2\pi}
\frac{J_0(\omega)\exp(-\mathi\omega\Lambda)}{2\cosh(\omega u)} \; ,
\nonumber\\[6pt]
\rho_0(k)&=& \frac{1}{2\pi} +\cos k
	       \int_{-\infty}^{\infty}\frac{\diff\omega}{2\pi}
               \frac{J_0(\omega)\exp(-\mathi\omega\sin k)}{1+\exp(2u|\omega|)}
\; ,
\end{eqnarray}
and $J_n(x)$ are $n$th-order Bessel functions.
With the help of~(\ref{eqn:TBA2satoshi}) we can transform the free-energy
density~(\ref{eqn:gibbs2}) into the form (see eq.~(4.21) of~\cite{ZHa})
\begin{eqnarray}
 f(T)&\equiv& C - T\alpha(T) -T\beta(T) \; ,\label{eqn:gibbs}\\[3pt]
 C &=&4\int_{-\infty}^{\infty}\diff \Lambda \sigma_0(\Lambda)
           {\rm Re}\left[\sqrt{1-(\Lambda-\mathi u)^2}\,\right]-u-\mu
= -e_0-\mu \;, 
\label{C} \\[3pt]
 \alpha(T)&=&\int_{-\pi}^{\pi}\diff k \rho_0(k) \ln(1+1/\zeta(k))\;, 
 \label{eqn:alpha}\\[3pt]
 \beta(T)&=&\int_{-\infty}^{\infty}\diff\Lambda
\sigma_0(\Lambda)\ln(1+\eta_1^{'}(\Lambda))\;.
 \label{eqn:beta}
\end{eqnarray}
With the help of~(\ref{eqn:gibbs}) the internal energy
\begin{equation}
 e(T,u)= f(T)-T\frac{\partial f(T)}{\partial T}
\end{equation}
can be expressed as
\begin{equation}
 e(T,u) = C + T^2\left(\frac{\partial \alpha(T)}{\partial T}
+\frac{\partial \beta(T)}{\partial T}\right) \; .
\label{eqn:internal-en}
\end{equation}

\section{The spin-disordered limit}
\label{sec:int}

\subsection{Definition of the limit}
\label{subsec:deflimit}

In the following 
%%%%%% begin Fabian March 18, 2006 %%%%%%%%%%
we consider the TBA equations in the regime
%%%%%% begin Fabian March 18, 2006 %%%%%%%%%%
\begin{equation}
 J \ll T \ll \Delta
\label{range}
\end{equation}
%%%%%% begin Fabian March 18, 2006 %%%%%%%%%%
for $B=0$ and $\mu=0$, i.e., the half-filled Hubbard chain in zero
magnetic field. Here,
%%%%%% begin Fabian March 18, 2006 %%%%%%%%%%
$J(U/t\to\infty)= 4t^2/U$ is the coupling strength
for the spin degrees of freedom, and $\Delta$ is the gap for charge
excitations, $\Delta(U/t\to\infty)=U-4t$.
%%%%%% begin Fabian March 18, 2006; Florian March 20%%%%%%%%%%
The inequalities (\ref{range}) imply that
\begin{equation}
\frac{1}{UT}\ll 1\quad ,\quad \exp\left(-\frac{U}{T}\right)\ll 1\; .
\end{equation}
%%%%%% end Fabian March 18, 2006; Florian March 20 %%%%%%%%%%
In this limit, the spin degrees of freedom are `hot', i.e., 
the charge degrees of freedom move in a random spin background.

\begin{figure}[ht]
\centerline{\includegraphics[width=8cm]{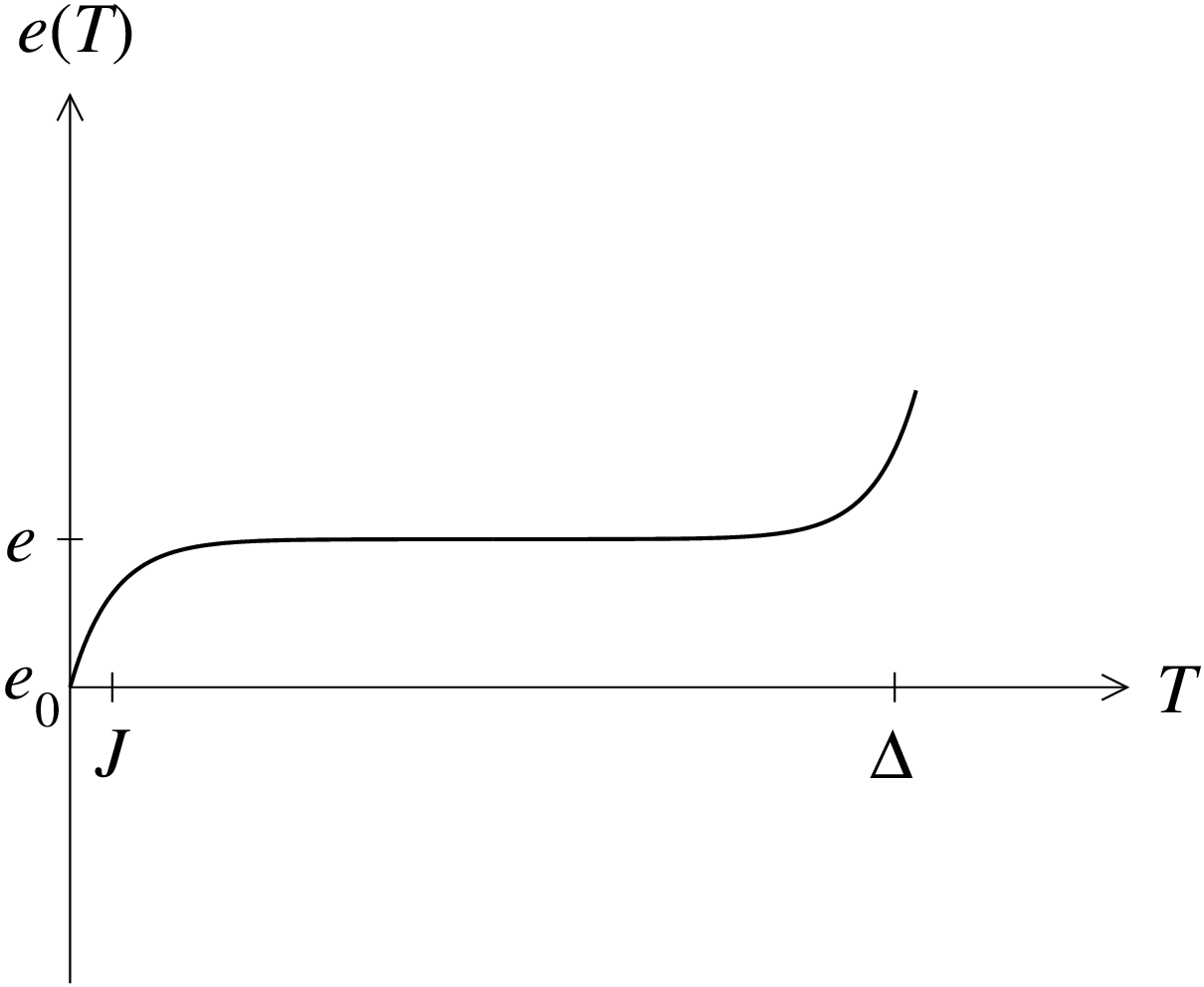}}
\caption{Internal energy as a function of temperature for the Hubbard model
at half band-filling and large interactions.\label{fig:internalenergy}}
\end{figure}

The qualitative dependence of the
internal energy as a function of temperature is shown 
in Fig.~\ref{fig:internalenergy}. In the temperature
range given by~(\ref{range}) the internal energy is essentially independent
of temperature because the spin degrees of freedom contribute maximally
to the internal energy whereas there are exponentially 
few charge excitations.

%%%%%%%%%%%%%%%%%%%%%%%%%%%%%%
\subsubsection{Charge sector}
%%%%%%%%%%%%%%%%%%%%%%%%%%%%%
We define the {\sl dressed energy\/} for charge excitations by
\begin{equation}
 \kappa(k, u, T)=T\ln(\zeta(k))\; .
\end{equation}
We will see that, in the spin-disordered regime, the dressed energy has
an expansion of the form
\begin{equation}
 \kappa(k, u, T)=
   \sum_{m=0}^{\infty}\left(\frac{1}{uT}\right)^m \kappa^{(m)}(k,u)
   +{\cal O}\bigl(e^{-u/T}\bigr)\; .
\label{eqn:expansion}
\end{equation}
The leading term in this expansion is temperature independent,
negative and of order ${\cal O}(u)$. Hence
\begin{equation}
 \zeta(k)={\cal O}\bigl(e^{-u/T}\bigr) \ll 1 \; ,
\label{eqn:zeta}
\end{equation}
\begin{equation}
 \ln(1+1/\zeta(k))= -\ln\bigl(\zeta(k)\bigr)+{\cal
 O}\bigl(e^{-u/T}\bigr)\; .
\label{eqn:zeta2}
\end{equation}
Equation~(\ref{eqn:zeta2}) leads to a significant simplification of the
TBA equations.
%%%%%%%%%%%%%%%%%%%%%%%%%%%%%%%%%%%%%%%%%%%%%%%%%
\subsubsection{$k$-$\Lambda$ Strings}
%%%%%%%%%%%%%%%%%%%%%%%%%%%%%%%%%%%%%%%%%%%%%%%%%
Neglecting terms of order ${\cal O}\bigl(e^{-u/T}\bigr)$, the
TBA equations for $k$-$\Lambda$ strings take the form
\begin{eqnarray}
\ln \eta'_n(\Lambda) &=& 
\left. s*\ln(1+\eta'_{n+1})(1+\eta'_{n-1})\right|_{\Lambda} \; ,
\label{TBAkl}
\end{eqnarray}
where we define $\eta'_0(\Lambda)=0$. The solution to (\ref{TBAkl})
under the boundary conditions (\ref{boundarycond}) with $\mu=0$ is
\cite{Takahashi7274} 
\begin{equation}
\eta'_n(\Lambda)= (n+1)^2-1\; .
\label{etacons}
\end{equation}
The corrections to $\eta'$ are of order
$\exp\Bigl(\exp(-u/T)\Bigr)$.

%%%%%%%%%%%%%%%%%%%%%%%%%%%%%%%%%%%%%%%%%%%%%%%%%
\subsubsection{$\Lambda$ Strings}
%%%%%%%%%%%%%%%%%%%%%%%%%%%%%%%%%%%%%%%%%%%%%%%%%
Neglecting terms of order ${\cal O}\bigl(e^{-u/T}\bigr)$, the
TBA equations for $\Lambda$ strings take the form
\begin{eqnarray}
\ln \eta_n(\Lambda) &=& 
\left. s*\ln(1+\eta_{n+1})(1+\eta_{n-1})\right|_{\Lambda} \; ,
\label{TBAl}
\end{eqnarray}
where we define 
\begin{equation}
\eta_0(\Lambda)=\exp\left(-\frac{4}{T}{\rm Re}\sqrt{1-\Lambda^2}\right)-1\; .
\end{equation}
The solution to (\ref{TBAl}) under the boundary conditions
(\ref{boundarycond}) with $B=0$ can be obtained by iterative
linearization, as shown in Ref.~\cite{ZHa}. The starting point is
the solution $\eta_n^{(0)}$ to the equations (\ref{TBAl}) with
$\eta_0=0$, i.e.,
\begin{equation}
\eta^{(0)}_n(\Lambda)= (n+1)^2-1\; .
\label{sol0}
\end{equation}
One then linearizes (\ref{TBAl}) around the solution (\ref{sol0}) by
writing $\eta_n=\eta_n^{(0)}+\eta_n^{(1)}$ (for $n\geq 1$) and keeping
only the terms linear in $\eta_n^{(1)}$. 
This gives the following set of linear integral equations
\begin{equation}
\frac{\eta_n^{(1)}}{\eta_n^{(0)}}
=s*\frac{\eta_{n-1}^{(1)}}{1+\eta_{n-1}^{(0)}}+ 
s*\frac{\eta_{n+1}^{(1)}}{1+\eta_{n+1}^{(0)}}\; .
\label{set1}
\end{equation}
The boundary conditions are given by (\ref{boundarycond}) and
\begin{equation}
\eta_0^{(1)}(\Lambda)=-\frac{4}{T}{\rm Re}\sqrt{1-\Lambda^2}\; .
\end{equation}
The set of linear integral equations~(\ref{set1}) can
be solved by Fourier transform~\cite{ZHa}. We find
\begin{eqnarray}
\eta_n^{(1)}(\Lambda)&=&
\frac{2(n+1)n}{T} 
{\rm Re} \left[\sqrt{1-\bigl(\Lambda-\mathi u(n+2)\bigr)^2}\right] 
\nonumber \\
&& 
-\frac{2(n+1)(n+2)}{T} 
{\rm Re} \left[\sqrt{1-(\Lambda-\mathi nu)^2}\right]  \nonumber \\
&=& {\cal O}\left( \frac{1}{uT} \right)
\; .
\end{eqnarray}
We see that we have
\begin{equation}
|\eta_1^{(1)}(\Lambda)|\ll \eta_1^{(0)}(\Lambda)=3 \; ,
\end{equation}
as required.

%%%%%%%%%%%%%%%%%%%%%%%%%%%%%%%%%%%%%%%%%%%%%%%%%%%%%%%%%%%%%%%%%%%
\subsection{Dressed  energy of the charge degrees of freedom} 
\label{subsec:cdgf}
%%%%%%%%%%%%%%%%%%%%%%%%%%%%%%%%%%%%%%%%%%%%%%%%%%%%%%%%%%%%%%%%%%%
Combining the results for $\eta_1(\Lambda)$ and $\eta'_1(\Lambda)$ we
obtain 
\begin{equation}
 \ln\left[\frac{1+\eta_1^{\prime}(\Lambda)}{1+\eta_1(\Lambda)}\right]
 \approx -\frac{1}{4}\eta_1^{(1)}(\Lambda)\; .
\end{equation}
Substituting this back into (\ref{eqn:TBA2satoshi}) we obtain the first term
of the expansion (\ref{eqn:expansion})
\begin{equation}
\kappa^{(0)}(k,u)=
-2\cos k -2u -\frac{1}{2}\int_{-\infty}^{\infty}\frac{d\omega}{\omega}
           J_1(\omega)e^{\mathi \omega\sin k}e^{-2u|\omega|} \; .
\label{eqn:kappa}
\end{equation}
Crucially, the contributions of $\eta_1$ and $\eta'_1$ to $\zeta(k)$
do not feed back into the integral equations for $\eta_n$ and
$\eta'_n$. The reasons for this are that 
\begin{enumerate}
\item the corrections do not change the fact that $\zeta$ 
is ${\cal O}(\exp(-u/T))$ 
and terms involving $\ln(1+\zeta(k))$ can therefore be
dropped from the TBA equations;
\item the corrections to $\ln(\zeta)$ depend only on $\sin(k)$ and
therefore do not contribute to
\begin{equation}
\int_{-\pi}^{\pi}\diff k \cos(k)\ s(\Lambda-\sin k)\ln[\zeta(k)]\; .
\end{equation}
\end{enumerate}
The result~(\ref{eqn:kappa}) should be compared to the corresponding
expression at zero temperature, see eq.~(7.10) of~\cite{Zebook},
\begin{equation}
\kappa_0(k,u)=-2\cos k -2u 
   -2 \int_{-\infty}^{\infty}\frac{d\omega}{\omega} J_1(\omega)
e^{\mathi \omega\sin k} \frac{1}{1+\exp(2 |\omega| u)} \; .
\label{eqn:kappazero}
\end{equation}
Performing the integral in~(\ref{eqn:kappa}) we finally find ($u=U/4$)
\begin{equation}
 \kappa^{(0)}(k,U)=\epsilon(k)-\frac{1}{2\sqrt{2}}\sqrt{[\epsilon(k)]^2+U^2
                   +\sqrt{[\epsilon(k)^2-U^2]^2+(4U)^2}} \; ,
\label{eqn:kappafinal}
\end{equation}
where $\epsilon(k)=-2\cos k$ is the bare dispersion.

The interesting point is that $\kappa^{(0)}(k)\neq \kappa_0(k)$
despite the fact that $T\ll |\kappa_0(k)|$. The reason for this is
that spin and charge degrees of freedom are still coupled in the
Hubbard model. This is obvious from the TBA equations and also
from the known scattering matrix of elementary 
excitations~\cite{EsslerKorepin}, see Chap.~7.4 in~\cite{Zebook}.
We note that the modification of the dressed energy
of the charge degrees of freedom by the spin sector
is a general characteristic of the spin-disordered regime.
In the asymptotic low-energy regime, spin and charge degrees of
freedom decouple quite generically for one-dimensional 
interacting-electron systems. However, on the scale of the temperature in
the spin-disordered regime such a decoupling no longer holds.
This in turn is expected to lead to modifications in
non-universal physical properties such as the charge velocity.
Our result is in agreement with this expectation which we note 
equally applies to a less than half-filled band.

%%%%%%%%%%%%%%%%%%%%%%%%%%%%%%%%%%%%%%%%%%%%%%%%%%%%%%%%%%%%%%%%%%%
\subsection{Dressed momentum of the charge degrees of freedom} 
\label{subsec:dressedmtm}
%%%%%%%%%%%%%%%%%%%%%%%%%%%%%%%%%%%%%%%%%%%%%%%%%%%%%%%%%%%%%%%%%%%
The total momentum in thermal equilibrium is equal to zero by virtue
of translational invariance. In order to determine the dressed 
momentum of our charge excitation, we start from the expression
for the contribution of a charge excitation with real $k$'s to the
total momentum~\cite{Woynarovich} (see~(5.97) and (5.38) of~\cite{Zebook})
\begin{equation}
p(k)=k+\sum_{n=1}^{\infty} \int_{-\infty}^{\infty} \diff\Lambda
\theta\left(\frac{\sin k -\Lambda}{nu}\right)
\left[ \sigma_n^{\prime p}(\Lambda)+\sigma_n^{p}(\Lambda) \right]
\; ,
\label{eqn:realk}
\end{equation}
where
\begin{equation}
\theta(x)= 2\arctan\left(x\right) 
\end{equation}
and the root densities for particles and holes obey 
(see~(5.41) of~\cite{Zebook})
\begin{eqnarray}
\rho^{p}(k) + \rho^{h}(k) &=&\frac{1}{2\pi}  +\cos k
\sum_{n=1}^{\infty} \int_{-\infty}^{\infty} \diff\Lambda
a_n(\sin k -\Lambda) 
\left[ \sigma_n^{\prime p}(\Lambda) +\sigma_n^{p}(\Lambda)\right]
\label{eqn:erstexx}
\, , \\
\sigma_n^{p}(\Lambda) &=& -\sigma_n^{h}(\Lambda)+ 
\left. s *\left(\sigma_{n+1}^{h}+\sigma_{n-1}^{h}\right)\right|_{\Lambda}
\nonumber \\
&& +\delta_{n,1}
\int_{-\pi}^{\pi} \diff k s(\Lambda-\sin k) \rho^{p}(k)
\label{eqn:zweitexx}\; , \\
\sigma_n^{\prime p}(\Lambda) &=& -\sigma_n^{\prime h}(\Lambda)+ 
\left. s *\left(\sigma_{n+1}^{\prime h}
+\sigma_{n-1}^{\prime h}\right)\right|_{\Lambda}
\nonumber \\
&& -\delta_{n,1}
\int_{-\pi}^{\pi} \diff k s(\Lambda-\sin k) 
\left(\rho^{p}(k)-\frac{1}{2\pi}\right)
\label{eqn:drittexx}
\end{eqnarray}
with
\begin{equation}
\eta_n(\Lambda)=\frac{\sigma_{n}^h(\Lambda)}{\sigma_{n}^p(\Lambda)} 
\quad , \quad 
\eta_n^{\prime}(\Lambda)=
\frac{\sigma_{n}^{\prime h}(\Lambda)}{\sigma_{n}^{\prime p}(\Lambda)} 
\quad , \quad 
\zeta(k)=\frac{\rho^h(k)}{\rho^{p}(k)} \; .
\end{equation}
As $\zeta(k)={\cal O}\left(\exp(-u/T)\right)$ we can drop
$\rho^{h}(k)$ from equation~(\ref{eqn:erstexx}) and substitute
the resulting equation into~(\ref{eqn:zweitexx}) and
(\ref{eqn:drittexx}). We obtain the following result for the driving term
\begin{equation}
\int_{-\pi}^{\pi} \diff k s(\Lambda-\sin k) \rho^{p}(k)
=
\int_{-\pi}^{\pi} \frac{\diff k}{2\pi}  s(\Lambda-\sin k) 
= \sigma_0(\Lambda)
\; .
\end{equation}
The temperature-independent contribution to the particle and hole root
densities is obtained by using~(\ref{etacons}) and ~(\ref{sol0})
and then solving the resulting sets of coupled linear integral
equations. We find 
\begin{equation}
\sigma_n^{\prime p,h}(\Lambda)=
{\cal O}\left(\exp(-u/T)\right)
\end{equation} 
and
\begin{eqnarray}
\sigma_n^{p}(\Lambda) &=& \int_{-\infty}^{\infty}
\frac{\diff\omega}{2\pi} 
\frac{J_0(\omega)\exp\left(-\mathi \omega \Lambda\right)}{2(n+1)}
\left[
\frac{\exp(-nu|\omega|)}{n} - \frac{\exp(-(n+2)u|\omega|)}{n+2}
\right] \nonumber \\[3pt]
&& + {\cal O}\left(\exp(-u/T)\right)\\
&=&- \frac{1}{2\pi(n+1)(n+2)} 
{\rm Re}\left(\frac{1}{\sqrt{1-(\Lambda-\mathi (n+2) u)^2\,}}\right)
\nonumber \\
&& + \frac{1}{2\pi n(n+1)} 
{\rm Re} \left( \frac{1}{\sqrt{1-(\Lambda-\mathi n u)^2\,}}\right)
+ {\cal O}\left(\exp(-u/T)\right)
\; .
\label{eqn:sigmafinal}
\end{eqnarray}
Substituting~(\ref{eqn:sigmafinal}) back into~(\ref{eqn:realk})
we obtain the temperature-independent contribution to the dressed
momentum
\begin{equation}
p(k)= k+\frac{1}{2}\arcsin\left[    
\frac{2\sin k}{\sqrt{4u^2+(\sin k +1)^2} +\sqrt{4u^2+(\sin k -1)^2} }
\right] \;.
\label{eqn:finalp}
\end{equation}
We note that the physical momenta of holons~($h$) and 
anti-holons~($\overline{h}$)
are obtained from $p(k)$ by $p_h(k)=\pi/2-p(k)=p_{\overline{h}}(k)+\pi$.

%%%%%%%%%%%%%%%%%%%%%%%%%%%%%%%%%
\subsection{Effective dispersion for strong coupling}
\label{subsec:effdisp}
%%%%%%%%%%%%%%%%%%%%%%%%%%%%%%%%%

The effective dispersion is given implicitly by
equations~(\ref{eqn:kappafinal}) and~(\ref{eqn:finalp}).
In order to have a consistent expansion for the
Hubbard model in the spin-disordered regime, we need to truncate
these two equations at order $1/u$ because the next subleading
temperature-dependent contributions are of the order $1/(uT)$.
Hence, the effective dispersion in the spin-disordered Hubbard model
is given by
\begin{equation}
\kappa^{(0)}(p,u)=-2u -2\cos p -\frac{1}{4u}(3-2 \cos^2 p) 
+ {\cal O}\left( u^{-2} \right) \; .
\label{eqn:disp}
\end{equation}
For the Hubbard model at zero temperature we find instead
\begin{equation}
\kappa_0(p,u)=-2u -2\cos p -\frac{\ln 2}{u}(3-2\cos^2 p)
+ {\cal O}\left( u^{-2} \right) \; .
\label{eqn:disphubbard}
\end{equation}
Both formulae can be cast into the form
\begin{equation}
\kappa(p,u,T)=-2u -2\cos p - \frac{1}{4u}
\left( 1-4 \gamma_{\rm s}(T) \right) 
(3-2\cos^2 p)
+ {\cal O}\left( u^{-2} \right) \,,
\label{eqn:NEW}
\end{equation}
where 
\begin{equation}
\gamma_{\rm s}(T)=\langle
\hat{\mbox{\bf S}}_i\hat{\mbox{\bf S}}_{i+1}
\rangle_{\rm s}
\end{equation}
denotes the nearest-neighbor spin correlation function in the 
Heisenberg model at temperature~$T$. Thermal averages of 
operators~$\hat{A}$ over spin configurations
are defined by
\begin{equation}
\langle \hat{A} \rangle_{\rm s} = 
\frac{{\rm Tr} \left[\exp\left(-\beta \hat{H}_{\rm Heis}\right) 
\hat{A}\right]}{
{\rm Tr} \left[\exp\left(-\beta \hat{H}_{\rm Heis}\right)\right]}
\end{equation}
with the Heisenberg Hamiltonian (see Appendix~2.A of~\cite{Zebook})
\begin{equation}
\hat{H}_{\rm Heis} = \sum_{i} \frac{4t^2}{U}
\left(\hat{\mbox{\bf S}}_i\hat{\mbox{\bf S}}_{i+1} -\frac{1}{4}\right)\; .
\label{eqn:newHeis}
\end{equation}
In fact, we have $\gamma_{\rm s}(T=0)=1/4-\ln(2)$ from the 
Bethe-Ansatz solution of the Heisenberg model~\cite{Bethe}
and $\gamma_{\rm s}(T=\infty)=0$ for uncorrelated spins.
%%%%%% begin Florian March 17, 2006 %%%%%%%%%%
Results for all temperatures can be found in Refs.~\cite{Japanerseries,Bortz}.
%%%%%% end Florian March 17, 2006 %%%%%%%%%%

The result~(\ref{eqn:NEW}) can be obtained within
the $1/U$-expansion~\cite{GebhardKalinowski}.
This approach can be used here
because the one-dimensional lattice is a Bethe lattice with
coordination number~$Z=2$. 
To leading order in~$1/u$ we must treat the Hamiltonian
\begin{equation}
\hat{h}_1 = -\frac{1}{U}\hat{P}_{0} \hat{T} \hat{P}_1 \hat{T} \hat{P}_{0}
\; , 
\end{equation}
where $\hat{P}_0$ projects onto the subspace of zero double occupancies.
At half band-filling, $\hat{h}_1$ reduces to the Heisenberg 
model~(\ref{eqn:newHeis}).
The internal energy density of the Heisenberg model is given by
\begin{equation}
e_{\rm s}(T) =\frac{1}{u}\left(\gamma_{\rm s}(T)-\frac{1}{4} \right) \;,
\label{eqn:tobedetermined}
\end{equation}
and can be calculated analytically in terms of the Thermodynamic
Bethe Ansatz~\cite{MTakahashi,Takahashi71} or, equivalently, 
via the solution of a set of coupled integral equations~\cite{Kluemper}.

For the derivation of the effective dispersion we have to
solve (see~(43) of~\cite{GebhardKalinowski})
\begin{eqnarray}
\frac{U}{L} \sum_{j,\sigma} \left\langle 
\hat{c}_{j,\sigma}^{+} 
\left(\hat{h}_1 -Le_{\rm s}(T)\right) 
\hat{c}_{j,\sigma}^{\phantom{+}}
\right\rangle_{\rm s}
&=& 
\frac{1}{L} \sum_{j,\sigma} \left\langle 
\hat{c}_{j,\sigma}^{+} 
\left[ g_{1,2} (\hat{h}_0)^2 +g_{1,0}\right] 
\hat{c}_{j,\sigma}^{\phantom{+}}
\right\rangle_{\rm s} \nonumber \\
&=&
2 g_{1,2} +g_{1,0} \; , \label{eqn:lhs1}
\end{eqnarray}
where $\hat{h}_0$ describes the the hopping of holes in the spin
background, $\hat{h}_0=\hat{P}_0\hat{T}\hat{P}_0$.
The expectation value on the left-hand-side of~(\ref{eqn:lhs1})
is readily calculated so that we find
\begin{equation}
-8\left(\gamma_{\rm s}(T)-\frac{1}{4} \right) =2g_{1,2} +g_{1,0}
\end{equation}
as our first equation.
The second equation we obtain from the solution
of eq.~(44) of~\cite{GebhardKalinowski},
{\arraycolsep=0pt\begin{eqnarray}
\frac{U}{L} \sum_{j,\sigma} \Bigl\langle 
\hat{c}_{j,\sigma}^{+} 
\left(\hat{h}_1 -Le_{\rm s}(T)\right) && (\hat{h}_0)^2 
\hat{c}_{j,\sigma}^{\phantom{+}}
\Bigr\rangle_{\rm s} 
 \nonumber \\
&& = \frac{1}{L} \sum_{j,\sigma} \left\langle 
\hat{c}_{j,\sigma}^{+} 
\left[ g_{1,2} (\hat{h}_0)^2 +g_{1,0}\right] (\hat{h}_0)^2 
\hat{c}_{j,\sigma}^{\phantom{+}}
\right\rangle_{\rm s} 
\nonumber \\
&&=
6g_{1,2} +2 g_{1,0} \; . \label{eqn:lhs2}
\end{eqnarray}}
The expectation value on the left-hand-side of~(\ref{eqn:lhs2})
is readily calculated and we find
\begin{equation}
4\left(\gamma_{\rm s}(T)-\frac{1}{4} \right) =2g_{1,2} 
\end{equation}
as our second equation. From this 
$g_{1,2}=-(1-4\gamma_{\rm s}(T))/2$ and $g_{1,0}=3(1-4\gamma_{\rm s}(T))$ 
result, and the effective Hamiltonian
for the motion of a single hole in a spin background becomes
\begin{equation}
\hat{h}^{\rm eff}= 
\hat{h}_0 +\left(1-4\gamma_{\rm s}(T)\right)
\frac{\left(-(\hat{h}_0)^2/2 + 3\right)}{U}
+{\cal O}\left(U^{-2}\right) \; .
\end{equation}
Replacing $\hat{h}_0\to -\epsilon(p)=2\cos p$ and $\omega+2u=
-\hat{h}^{\rm eff}$ as in~\cite{GebhardKalinowski} we find for 
$\kappa(p,u,T)\equiv \omega$
\begin{equation}
\kappa(p,u,T)=-2u -2\cos p -\frac{1}{4u}
\left( 1-4 \gamma_{\rm s}(T) \right) 
(3-2 \cos^2 p) 
+ {\cal O}\left( u^{-2} \right) \, ,
\label{eqn:displargeUexp}
\end{equation}
as used in~(\ref{eqn:NEW}). 

{}From the $1/U$-expansion we can determine the density of
states of the lower Hubbard band. As in~\cite{GebhardKalinowski} the
shape-correction factor to first order is found to be
$s(\epsilon)=1-(1-4\gamma_{\rm s}(T))\epsilon/U$ so that we find
[$\alpha(T)=1-4 \gamma_{\rm s}(T)$]
\begin{eqnarray}
D_{\rm LHB}^{(1)}(\omega) &=& \int_{-2}^{2}\diff\epsilon
\rho_0(\epsilon) \left(1-\alpha(T)\frac{\epsilon}{U}\right)
\delta\left(\omega+U/2+\epsilon +\alpha(T)\frac{6-\epsilon^2}{2U}\right)
\nonumber \\[3pt]
&=&\rho_0\left[\left(
U-\sqrt{(1+\alpha(T))U^2+2\alpha(T)U\omega+6\alpha(T)^2}
\right)/\alpha(T)
\right]
\, ,
\label{eqn:DOS}
\end{eqnarray}
where $\rho_0(\epsilon)=1/(\pi\sqrt{4-\epsilon^2})$ for $|\epsilon|<2$ is the 
density of states for non-interacting electrons and
$\omega_{-}<\omega<\omega_{+}$ with $\omega_{\pm}=-U/2\pm 2-\alpha(T)/U$.
In particular, for the single-particle gap we find
\begin{equation}
\Delta^{(1)}(u,T)=-2\omega_{+} = 4u-4+\frac{1-4 \gamma_{\rm s}(T)}{2u} \; ,
\label{eqn:tobedetermined-delta}
\end{equation}
up to and including the first order in the strong-coupling expansion.

Finally, we note that the momentum distribution can also be determined
along these lines. We find 
\begin{equation}
\langle \hat{n}_{p}\rangle_{\rm s} = 
\sum_{\sigma} \langle \hat{n}_{p\sigma}\rangle_{\rm s} = 
1- \frac{\cos p}{2u}\left( 1-4 \gamma_{\rm s}(T) \right) 
+ {\cal O}\left( u^{-2} \right) \; ,
\end{equation}
in agreement with eq.~(3.2) of Ref.~\cite{Takahashi77}.

%%%%%%%%%%%%%%%%%%%%%%%%%%%%%%%%%
\subsection{Internal energy}
\label{subsec:internalenergy}
%%%%%%%%%%%%%%%%%%%%%%%%%%%%%%%%%
We must evaluate~(\ref{eqn:internal-en}) in the spin-disorder limit.
We have
\begin{eqnarray}
 e_{\alpha}&\equiv&  T^2\frac{\partial \alpha}{\partial T}
\approx \int_{-\pi}^{\pi}\diff k \rho_0(k) \kappa^{(0)}(k,u)\;, 
\label{eqn:ealpha}
\\
 e_{\beta}&\equiv& T^2\frac{\partial \beta}{\partial T}
=0\; .
\end{eqnarray}
The latter follows from the fact that $\eta_1'$ is independent of 
temperature. This leads to the following expression for the internal energy
\begin{eqnarray}
 e(T,u)&=&-u-\int_{0}^{\infty}\frac{\diff\omega}{\omega}
      J_0(\omega)J_1(\omega) \exp\left(-2u\omega\right)+{\cal
      O}(1/Tu)\nonumber\\
&=&- \frac{2u}{\pi}{\rm E}\left(-\frac{1}{u^2}\right)+{\cal O}(1/Tu)
\equiv e(u)+{\cal O}(1/Tu) \; .
\label{eqn:eint}
\end{eqnarray}
{}From the internal energy we can derive the average
double occupancy, $d(u)=(1/4)(\partial [e(T,u)+u])/(\partial u)$, as
\begin{equation}
 d(u)=\frac{1}{4}-\frac{1}{2\pi}
       {\rm K}\left(-\frac{1}{u^2}\right)+{\cal O}(1/Tu) \; .
\label{eqn:double}
\end{equation}
Here ${\rm K}(m)$ and ${\rm E}(m)$ are the complete
elliptic integrals of the first and second kind, respectively.
The internal energy is to be compared to the ground-state energy of
the Hubbard model at half band-filling
\begin{equation}
 e_0(u)=-u-4\int_{0}^{\infty}\frac{\diff\omega}{\omega}
      J_0(\omega)J_1(\omega) \frac{1}{1+\exp\left(2u\omega\right)} \;.
\label{eqn:ezerozero}
\end{equation}
The average double occupancy for the Hubbard model at zero temperature 
follows from the derivative of~(\ref{eqn:ezerozero}) with respect to
the interaction strength.

\section{Zero-temperature interpretation}
\label{sec:interpretation}

\subsection{Mott--Hubbard transition}

Now we {\sl interpret\/} our results in terms of a putative one-dimensional
interacting-electron system at zero temperature. In practice, we use
the results for the temperature-independent contributions to the
internal energy and the effective dispersion derived for the
spin-disorder regime in the Hubbard model, $J\ll T \ll \Delta$,
for {\sl any\/} value of~$U$. Our motivation are studies of the
Hubbard model in infinite dimensions where the Mott--Hubbard insulator
is in the spin-disordered phase above the Mott--Hubbard transition.
In order to model such a situation in a one-dimensional system
one would need to take the limit $T\to 0$ {\sl after\/} 
letting $J\to 0$. Of course, this is not possible for
the Hubbard model. 

The one-dimensional Hubbard model at half band-filling describes 
a Mott--Hubbard insulator for all $U>0$, i.e., 
the gap~$\Delta_0(U)/2$ for single-particle charge excitations is finite,
$\Delta_0(U) = -2\kappa_0(\pm\pi,U)$. From~(\ref{eqn:kappazero}) we have
\begin{equation}
\Delta_0(U) = 2\left[-2+\frac{U}{2}+4\int_{0}^{\infty} \frac{\diff\omega}{\omega}
J_1(\omega) \frac{1}{1+\exp\left(U\omega/2\right)}\right] \; .
\end{equation}
For $U\to 0$, the gap is exponentially small whereas it increases
linearly with $U$ for large interactions.
The transition at $U_{\rm c}=0^+$ is readily understood as the consequence
of the perfect-nesting property in one dimension so that
the (marginally) relevant Umklapp scattering processes 
drive the system into the insulating phase for all $U>0$.

\begin{figure}[ht]
\centerline{\includegraphics[width=12cm]{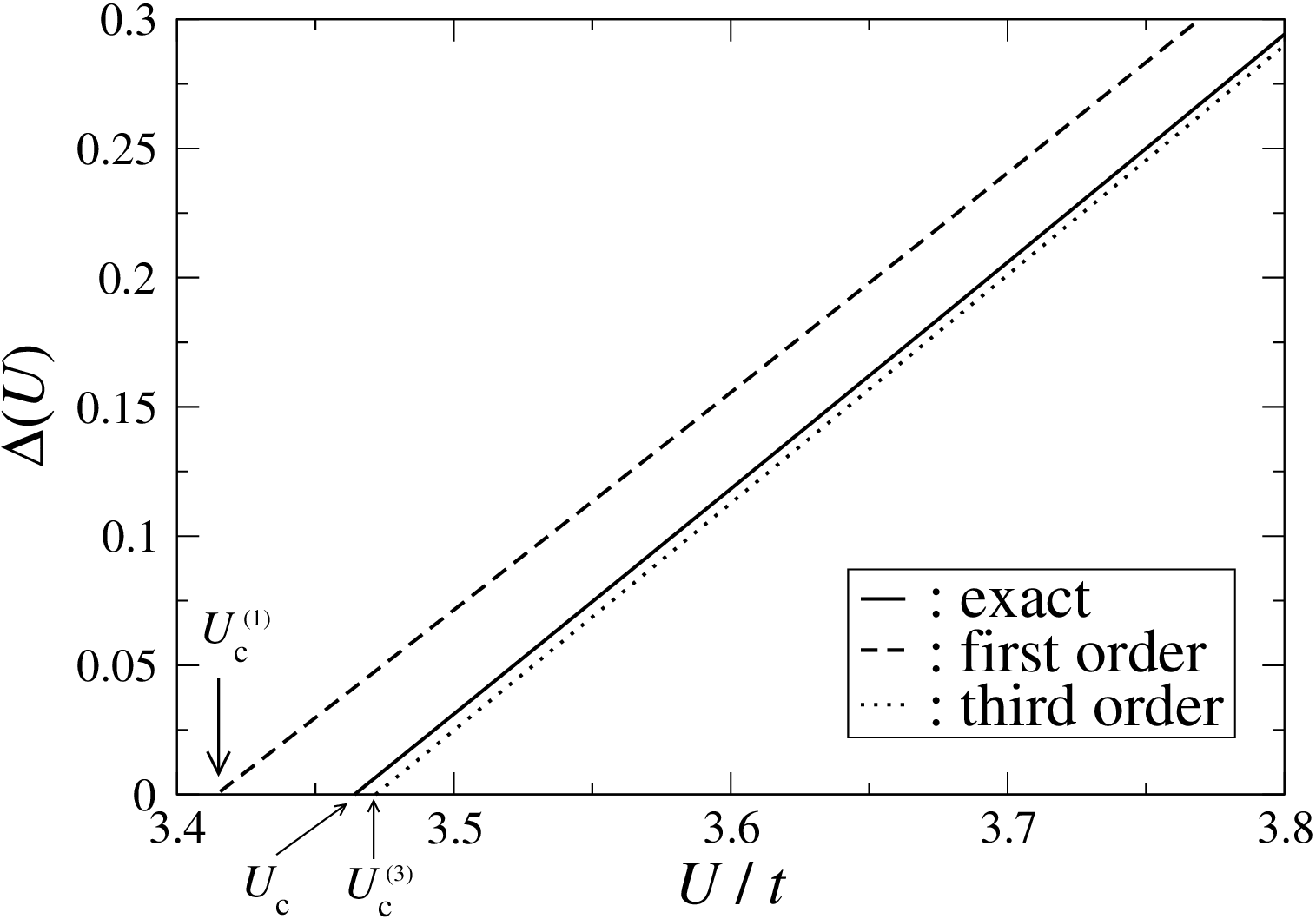}}
\caption{Gap for single-particle excitations as a function of 
the interaction strength~$U$ for the interacting-electron
system with a disordered spin background at half band-filling.\label{fig:gap}}
\end{figure}

In our putative interacting-electron system
these scattering processes are rendered ineffective by
the random spin background which, at zero energy cost,
provides a mechanism to dissipate momentum
in scattering processes involving charge degrees of freedom.
Therefore, we expect that the charge gap will open at a critical
interaction strength. Indeed, from~(\ref{eqn:kappa})
and~(\ref{eqn:kappafinal}) we find
\begin{equation}
\Delta(U) = 2\left[-2+\frac{U}{2}+\int_{0}^{\infty} \frac{\diff\omega}{\omega}
J_1(\omega) \exp\left(-U\omega/2\right)\right] 
= -4+\sqrt{U^2+4} \; .
\end{equation}
The gap opens linearly with slope $\sqrt{3}/2$ at 
\begin{equation}
U_{\rm c}=2\sqrt{3}=\frac{\sqrt{3}}{2}W\approx 0.866W\; ,
\end{equation}
where $W=4$ is the bandwidth of the Hubbard model.
The gap is shown in Fig.~\ref{fig:gap}.

\subsection{Average double occupancy}

\begin{figure}[ht]
\centerline{\includegraphics[width=12cm]{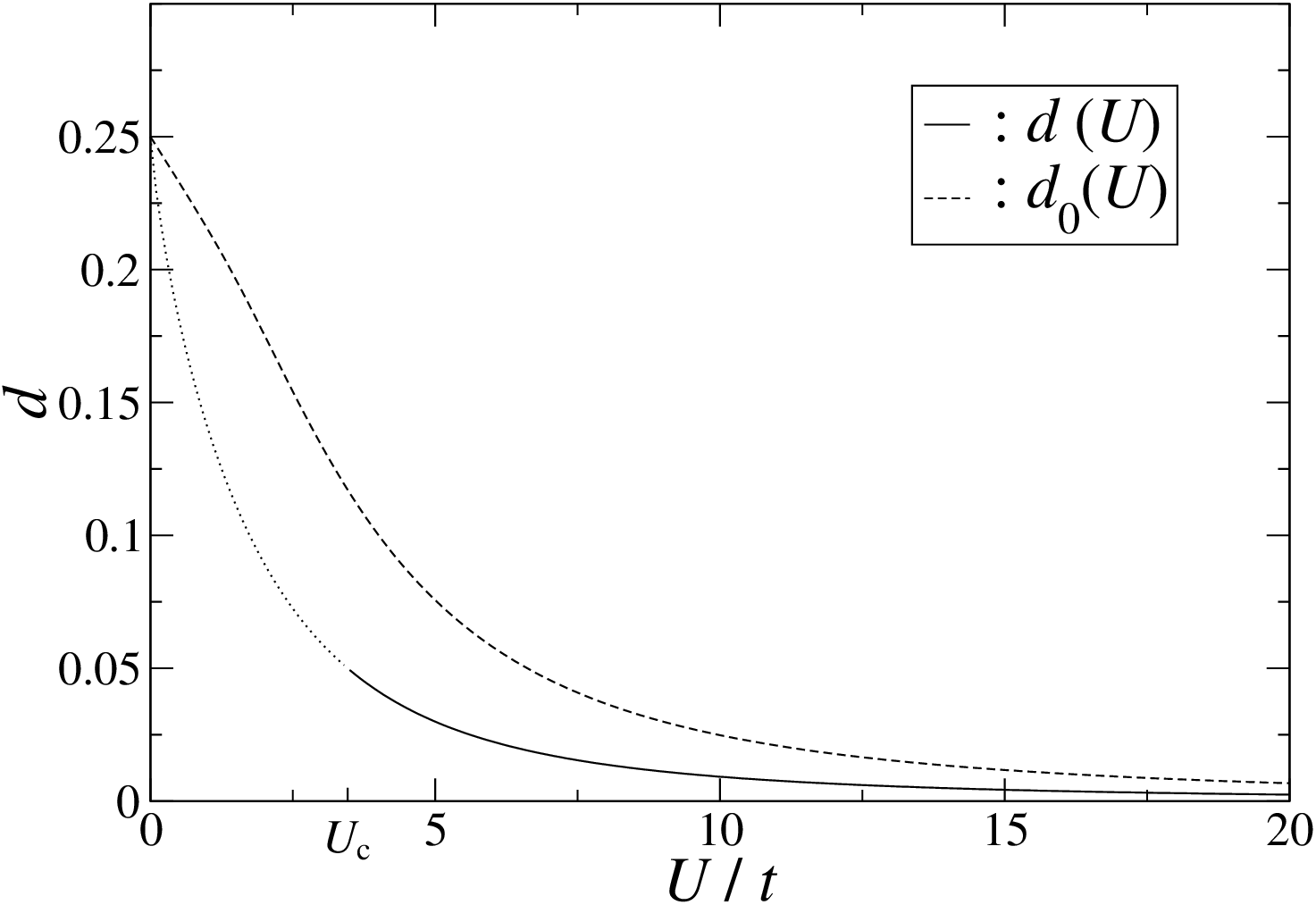}}
\caption{Average double occupancy~$d_0(U)$ of the Hubbard model
at half band-filling as compared to~$d(U)$~(\protect\ref{eqn:double}) 
for the interacting-electron system with a
spin-disordered background.\label{fig:compdouble}}
\end{figure}

We expect that physical quantities such as the average double
occupancy display unphysical behaviour in the region $U<U_{\rm c}$.
In fact, $d(u)$ contains a term
proportional to $u\ln(u)$ so that its derivative diverges
logarithmically for $u\to 0$. 
This diverging slope is seen in Fig.~\ref{fig:compdouble} where
we compare the average double occupancy
of the interacting-electron system with a spin-disordered background
with the double occupancy of the Hubbard model.

\subsection{Strong-coupling expansions}

The internal energy of the Hubbard model at zero temperature
and in the spin-disordered case can be expanded in powers of $1/U$.
In both cases, the radius of convergence is given by
$U_{\rm R}^{\rm e}=4$, see (6.83) of~\cite{Zebook} for $e_0(u)$ 
and (17.3.12) of~\cite{AS} for $e(u)$.
Explicitly,
\begin{eqnarray}
[e(U)+U/4]/U& = & \sum_{m=1}^{\infty} a_{2m} U^{-2m} \nonumber \\
&=&-\frac{1}{U^2}+\frac{3}{U^4}-\frac{20}{U^6}+\frac{175}{U^8}
          -\frac{1764}{U^{10}}+\frac{19404}{U^{12}} \pm \cdots \; .
\label{eqn:Uenergyexpand}
\end{eqnarray}
In~(\ref{eqn:tobedetermined}) we determined the coefficient~$a_2=-1$ 
within the $1/U$-expansion. From the coefficients~$a_{2m}$, 
one may actually deduce the radius of convergence
of the series, as done in~\cite{Bluemer}, by extrapolating
the ratio of the coefficients $r(m)=|a_{2m}/a_{2m-2}|$ ($m\geq 2$)
for $m\to\infty$. In our case $r(m)=4(2m-1)(2m-3)/m^2$ is a second-order 
polynomial in $1/m$ and we correctly find
\begin{equation}
U_{\rm R}^{\rm e}=\lim_{m\to\infty}\left(\sqrt{r(m)}\right)=4 \; .
\end{equation}
Note, however, that the radius of convergence of the energy
is not related to the critical interaction strength of the
metal-insulator transition, $U_{\rm c}=2\sqrt{3}$.
Therefore, it may also be doubted that this approach~\cite{Bluemer} 
is justified for the case of the infinite-dimensional Hubbard model.

In contrast to the internal energy, the series expansion of the gap 
in the spin-disordered Hubbard model converges for 
$U>U_{\rm R}^{\rm gap}=2$. The first terms of the expansion read
\begin{equation}
\Delta(U)=\sum_{m=-1}^{\infty}b_{m}U^{-m}
=U-4+\frac{2}{U}-\frac{2}{U^3}\pm \cdots \; .
\end{equation}
In~(\ref{eqn:tobedetermined-delta}) we determined
the coefficients~$b_{-1}=1$, $b_{0}=4$, and $b_{1}=2$
within the $1/U$-expansion.

Now that $U_{\rm c}=2\sqrt{3}$ is larger than the convergence radius of
the series, $U_{\rm R}^{\rm gap}=2$,
the gap opens linearly and 
the first few orders of the $1/U$-expansion provide a good
description of the gap, as shown in Fig.~\ref{fig:gap}.
This is particularly true for the critical values for the closing of the gap
as inferred from the truncated $1/U$-expansion.
Let us denote by $\Delta^{(m)}(U)$ the $m$th-order truncation of the series,
e.g., 
%%%%%% begin Florian March 17, 2006 %%%%%%%%%%
\begin{eqnarray}
\Delta^{(0)}(U)&=&U-4 \;, \nonumber \\
\Delta^{(1)}(U)&=&U-4-\frac{2}{U}\; , \nonumber \\ 
\Delta^{(3)}(U)&=&U-4-\frac{2}{U}-\frac{2}{U^3}\; , 
\end{eqnarray}
%%%%%% end Florian March 17, 2006 %%%%%%%%%%
etc., and let $U_{\rm c}^{(m)}$ be the critical interaction strength 
at which the $m$th gap opens, $\Delta^{(m)}(U_{\rm c}^{(m)})=0$.
We then find
\begin{eqnarray}
 U_{\rm c}^{(0)}&=& 4\; ,   \nonumber \\
 U_{\rm c}^{(1)}&=& 3.4142[1.4\%]\; ,   \nonumber \\
 U_{\rm c}^{(3)}&=& 3.4717[0.2\%]\; ,   \nonumber \\
 \vdots \nonumber \\
 U_{\rm c}\ &=&2\sqrt{3}=3.4651\; . 
\end{eqnarray}
The numbers in square brackets give the percentage difference to $U_{\rm c}$.
It is seen that the series converges very fast to the exact value.
This observation supports the application of this approach to
the Hubbard model in infinite dimensions~\cite{Nishimoto,GebhardKalinowski}.

\section{Conclusions}
\label{sec:conclusions}

We have analyzed the Thermodynamic Bethe Ansatz equations
for the half-filled one-dimensional Hubbard model in the
spin-disordered regime, $ t^2/U \ll T \ll U$.
We have derived explicit expressions for the leading terms
in the internal energy and the dressed energy and momentum
of the charge degrees of freedom. 
The resulting effective dispersion of holons and anti-holons differs
from the corresponding result in the half-filled Hubbard model
at zero temperature to order~$J=4t^2/U$.
This effect is due to the coupling of charge and spin degrees of
freedom and occurs at the expected energy scale.

We then interpreted the entire temperature-independent part
of the effective dispersion and the internal energy in terms
of a putative interacting-electron system at zero temperature. From 
these results we derived some implications for the analysis
of large-coupling expansions for the Mott--Hubbard insulator
in infinite dimensions. Moreover, our analytical result can be used
to assess the quality of other approximate schemes which 
describe the effective charge dispersion in a random spin background,
see, e.g., Ref.~\cite{Uhrig2}.

%%%%%%%%% Acknowledgment %%%%%%%%%%%%%%%%%%%%%%%%%%%%%%%%%%%%%%%
\ack
We are grateful to S.~Nishimoto for bringing Ref.~\cite{ZHa}
to our attention. SE is supported
by the Honjo International Scholarship Foundation.


\section*{References}

\begin{thebibliography}{99}

\bibitem{Mott} N F Mott, {\sl Metal--Insulator Transitions}, 2nd edition
(Taylor and Francis, London, 1990).

\bibitem{Gebhardbuch} F Gebhard, {\sl The Mott Metal-Insulator Transition}
(Springer, Berlin, 1997).

\bibitem{Metzner} W Metzner and D Vollhardt, Phys.~Rev.~Lett.~{\bf 62}, 324
(1989).

\bibitem{RMP} A Georges, G Kotliar, W Krauth, and
M J Rozenberg, Rev.\ Mod.\ Phys.~{\bf 68}, 13 (1996).

\bibitem{Nishimoto} S Nishimoto, F Gebhard, and E Jeckelmann, 
J.\ Phys.\ Cond.\ Matt.~{\bf 16}, 7063 (2004).

\bibitem{Uhrig} M Karski, C Raas, and G S Uhrig,
Phys.~Rev.~B~{\bf 72}, 113110 (2005).

\bibitem{GebhardKalinowski} M P Eastwood, F Gebhard, E Kalinowski, S Nishimoto,
and R M Noack, Eur.\ Phys.\ J.\ B~{\bf 35}, 155 (2003).

\bibitem{Bluemer} N Bl\"umer and E Kalinowski,
Phys.\ Rev.\ B~{\bf 71}, 195102 (2005).

\bibitem{FKreview} J K Freericks and V Zlatic,
Rev.\ Mod.\ Phys.~{\bf 75}, 1333 (2003).

\bibitem{Zebook} F H L Essler, H Frahm, F G\"ohmann, A Kl\"umper, and
V E Korepin, {\sl The one-dimensional Hubbard model\/}
(Cambridge University Press, Cambridge, 2005).

\bibitem{Marburger} F Gebhard, K Bott, M Scheidler, P Thomas, and S W Koch, 
Phil.\ Mag.\ B~{\bf 75}, 13 (1997).

\bibitem{ChZ} V V Cheianov and M B Zvonarev, 
Phys.~Rev.~Lett.~{\bf 92}, 176401 (2004).

\bibitem{Balents} G A Fiete and L Balents, 
Phys.~Rev.~Lett.~{\bf 93}, 226401 (2004).

\bibitem{Takahashi7274} M Takahashi, Prog.~Theor.~Phys.~{\bf 47},
69 (1972); Prog.~Theor.~Phys.~{\bf 52}, 103 (1974).

\bibitem{MTakahashi} M Takahashi, 
{\sl Thermodynamics of One-Dimensional Solvable Models\/}
(Cambridge University Press, Cambridge, 1999).

\bibitem{ZHa} Z N C Ha,
Phys.\ Rev.\ B~{\bf 46}, 12205 (1992).

\bibitem{EsslerKorepin} F H L Essler and V E Korepin,
Phys.~Rev.~Lett.~{\bf 72}, 908 (1994); Nucl.~Phys.~B~{\bf 426},
505 (1994).

\bibitem{Woynarovich} F Woynarovich, J.~Phys.~C~{\bf 15}, 97 (1982);
{\sl ibid.}, 6397 (1982).

\bibitem{Bethe} H Bethe, Z.~Phys.~{\bf 71}, 205 (1931); see also eq.~(6.83)
of~\cite{Zebook}.

\bibitem{Japanerseries} Z Tsuboi and M Shiroishi, J.~Phys.~A~{\bf 38},
L363 (2005).

\bibitem{Bortz} M Bortz and F G\"ohmann, Eur.~Phys.~J.~B~{\bf 46}, 399 (2005).


\bibitem{Takahashi71} M Takahashi, Prog.~Theor.~Phys.~{\bf 46},
401 (1971).

\bibitem{Kluemper} A Kl\"umper, Z.~Phys.~{\bf 91}, 507 (1993).

\bibitem{Takahashi77} M Takahashi, J.~Phys.~C~{\bf 10}, 1289 (1977).


\bibitem{AS} M Abramovitz and I A Stegun, 
{\sl Handbook of Mathematical Functions} (Dover, New York, 1972).

\bibitem{Uhrig2} A Reischl, E M\"uller-Hartmann, and G S Uhrig,
Phys.~Rev.~B~{\bf 70}, 245124 (2004).

\end{thebibliography}
\end{document}